# On Remote Phonon Scattering

A. Dyson[1] and B. K. Ridley[2]
(1) School of Maths, Stats & Physics, Newcastle University, UK
(2) School of Computing Science and Electronic Engineering, University of Essex, Colchester, UK

**Abstract**
Polar phonons can induce electric fields in an adjacent layer, whether non-polar or polar, producing remote phonon scattering of electrons. Treatment of remote phonon scattering has been based on the dielectric continuum model which takes only the electrical boundary conditions into account. We show that crystals whose polar modes satisfy both mechanical and electric boundary conditions cannot introduce remote phonon effects in the absence of dispersion. Further, even in the presence of dispersion, remote phonon effects are negligible, as a consequence of the necessity of satisfying mechanical boundary conditions.

## 1. Introduction

The scattering of electrons in a non-polar layer, caused by electric fields associated with polar optical phonons in an adjacent polar layer, was treated by Wang and Mahan [1], and their result has been applied to the case of the silicon transistor [2]. Gate stacks including high-κ dielectrics have consistently shown a lower mobility than those with the native oxide [3]. The origin of the lower mobility was widely attributed to remote phonon scattering derived from the large ionic polarization of the high-κ material [2,4]. Remote Coulomb scattering has also been identified as a contributor to the lower mobility [5,6]. However, the concentration of Coulomb scattering centres required to reproduce the mobility reduction would result in a large threshold shift which is not observed in experiments [7]. Similar debates are present for other material systems making use of high-κ dielectrics e.g. InSe, $MoS_2$ [8]. Here we shown that remote phonon scattering is not responsible for mobility degradation because so few modes are able to satisfy the boundary conditions that remote phonon scattering is negligible.

The treatment of remote phonon or soft optical phonons up to now has been based on the dielectric continuum (DC) model of the interface mode. It is well-known that the DC model ignores the mechanical boundary conditions that must be obtained at the interface and, moreover, assumes a zero-dispersion model of the relevant modes. We point out here that if mechanical boundary conditions are taken into account, along with the usual electrical boundary conditions, the electric fields responsible for remote phonon scattering do not exist unless dispersion is also taken into account, and even then, the effect turns out to be negligible. In short, the DC model makes predictions that are not accurate. Dispersion is also a factor in the case of a polar barrier layer adjacent to a polar well-layer, but the same conclusion applies.

The requirement that both mechanical and electrical boundary conditions need to be satisfied for polar optical modes is well-known [9-11]. Meeting this requirement leads to normal modes of lattice vibration that are hybrids of longitudinally polarized (LO), transversely polarized (TO) and electromagnetic interface (IF) optical modes. In most cases, the difference in frequency of the adjacent layers is large enough for it to be assumed that the particle displacement at the interface must vanish i.e. **u**=0. Moreover, the disparity of the LO and TO frequencies means that a hybrid at the LO frequency must contain a TO mode with a large wave vector lying in the complex region of the dispersion, which makes the mode markedly evanescent. Being closely confined to the interface, the TO component serves to ensure that the component of displacement parallel to the plane of the interface is zero, but has negligible role otherwise. The triple hybrid can then be reduced to a double hybrid of LO plus IF whose properties must be such as to ensure that the normal component of the displacement vanishes at the interface. In this model, the mechanical boundary condition reduces effectively to $u_z=0$, where z is the coordinate perpendicular to the interface [12,13].

It is usually assumed for simplicity that the crystal is elastically isotropic, for then LO and TO modes become distinct. Also, the x coordinate can be taken to have arbitrary direction in the plane of the interface. We will further assume that the dimensions of the layers are large enough for the effect of distant surfaces and interfaces to be negligible, that is, we focus on the case of a single heterostructure.

**2. Polar/Non-Polar Case**

We take z=0 to define the interface, z≥0 to define the non-polar layer, and z≤0 to define the polar layer.

It is straightforward to obtain the components of the displacement **u** and the electric potential $\phi$ in the polar layer that satisfy the electrical and mechanical boundary conditions:

$$u_x = q_x A e^{i(q_x x - \omega t)} \left( \cos q_z z + \Gamma \sin q_z z - p e^{q_x z} \right)(1 - \delta_{z,0})$$
$$u_z = i q_z A e^{i(q_x x - \omega t)} \left( \sin q_z z - \Gamma \cos q_z z + \Gamma e^{q_x z} \right) \quad\quad 1$$
$$\phi = -i \alpha_B A e^{i(q_x x - \omega t)} \left( \cos q_z z + \Gamma \sin q_z z - s p e^{q_x z} \right)$$

$q_{x,z}$ are the components of the wave vector, and z≤0.

$$\Gamma = \frac{q_x}{q_z} p, \; p = \frac{1}{s(1+r)}, \; r = \frac{\varepsilon_{IF}}{\varepsilon_{NP}} \quad\quad 2$$

The relation between electric field **E** and displacement **u** is:

$$\mathbf{E} = -\alpha s \mathbf{u}, \; \alpha = \frac{e^*}{V_0 \varepsilon_\infty}, \; s = \frac{\omega^2 - \omega_T^2}{\omega_L^2 - \omega_T^2}, \; e^{*2} = M V_0 \omega_T^2 (\varepsilon_s - \varepsilon_\infty) \quad\quad 3$$

$V_0$ is the volume of the unit cell, M is the reduced mass, $\omega_{L,T}$ are the zone-centre frequencies of the LO and TO modes, $\varepsilon_s, \varepsilon_\infty$ are the static and high-frequency permittivities, and $\varepsilon_{IF}, \varepsilon_{NP}$ are the permittivities associated with the IF mode and the non-polar material. The permittivities associated with the LO and IF modes are:

$$\varepsilon_{LO} = \varepsilon_\infty \frac{\omega^2 - \omega_L^2 + v_L^2(q_x^2 + q_z^2)}{\omega^2 - \omega_T^2 + v_T^2(q_x^2 + q_z^2)}$$
$$\varepsilon_{IF} = \varepsilon_\infty \frac{\omega^2 - \omega_L^2}{\omega^2 - \omega_T^2} \quad\quad 4$$

We have assumed that the dispersion is quadratic in wave vector, with v the appropriate dispersion velocity. We note that $\varepsilon_{LO} = 0$, independent of wave vector.

The scattering potential in the non-polar material (z≥0) is:

$$\phi = -i\alpha_0 A \frac{r}{1+r} e^{i(q_x x - \omega t)} e^{-q_x z} \qquad 5$$

The amplitude is determined by energy normalization in operator form:

$$A = \frac{1}{Q}\left(\frac{\hbar}{2M\omega}\right)^{1/2}(a + a^\dagger)$$
$$Q^2 = \frac{1}{2}(q_x^2 + q_z^2)(1 + \Gamma^2) \qquad 6$$

We note that in the absence of dispersion the frequency of the hybrid is $\omega = \omega_L$ and $\varepsilon_{IF} = 0$; hence r=0. Eq.5 shows that in this case the scattering potential vanishes. This is always true of the LO component, for which $\varepsilon(\omega) = 0$, independent of dispersion; thus any remote-phonon effect is solely associated with the IF component. Dispersion is essential for there to be a remote-phonon effect, as this allows the IF frequency to equal that of the LO component, while keeping the associated permittivity finite.

At sufficiently high wave vector, dispersion makes $\varepsilon_{IF} = -\varepsilon_{NP}$, in which case r=-1, $p \to \infty$, and the IF mode dominates the interaction, with the possibility of providing a strong remote-phonon effect. This occurs when the hybrid frequency is:

$$\omega^2 = \frac{\varepsilon_\infty \omega_L^2 + \varepsilon_{NP} \omega_T^2}{\varepsilon_\infty + \varepsilon_{NP}} \qquad 7$$

This condition is exactly the condition assumed in the DC model. In this case there is a finite scattering potential. The scattering rate is then proportional to:

$$(e\phi)^2 = \frac{e^2 \hbar \omega}{V_0} \frac{(q_z/q_x)^2}{(q_z^2+q_x^2)} \left(\frac{1}{\varepsilon_\infty + \varepsilon_{NP}} - \frac{1}{\varepsilon_s + \varepsilon_{NP}}\right) \qquad 8$$

Apart from the wave-vector factor, which arises from energy normalization in a thick layer, this agrees with the result given by the DC model [1].

However, this approximation is only valid for large wave vectors. The calculation of the true scattering rate for the hybrid involves a sum over all of the optical modes in the lattice, their number determined by the number of unit cells in the lattice. The sum over modes is limited by the necessity of conserving energy and momentum, which limits the in-plane component of the wave vector, $q_x$, to be of order of the corresponding electron wave vector. A more severe limitation is the necessity for the IF frequency to satisfy the condition $\varepsilon_{IF}(\omega) + \varepsilon_{NP} = 0$. This limitation effectively rules out any effective remote-phonon effect. The situation is same for polar/polar structures.

Equation 9 shows the integration over $q_z$ which must satisfy the disappearance of the sum of the dielectrics. An insignificant number of modes have the appropriate frequency to satisfy this condition.

$$\int_{-\infty}^{\infty} dq_z \frac{L}{2\pi} \frac{(\cos q_z z - \Gamma \sin q_z z - spe^{-q_x z})^2}{(q_x^2 + q_z^2)(1+\Gamma^2)} \qquad 9$$

Figure 1 shows the integrand plotted as a function of frequency from $\omega_T$ to $\omega_L$ for a typical AlGaN/ GaN structure. A single mode is able to satisfy the disappearance of the sum of the dielectrics.

## 3. Conclusion

The necessity for polar modes to obey mechanical and electrical boundary conditions leads to a triple hybridization of LO, TO and IF modes. The widely employed dielectric continuum model makes predictions that are inaccurate due in part to its zero-dispersion assumption. The elastic properties of the crystal ensures that dispersion cannot be neglected. Indeed, dispersion has been shown to be directly responsible for the remote-phonon effect, but it turns out that so few modes can satisfy the conditions that the effect is negligible. We conclude that remote phonon scattering can not reduce mobility.

**Acknowledgements**
We are grateful to the US Office of Naval Research for support of this work under grant No. N000141812463 sponsored by Dr Paul Maki.

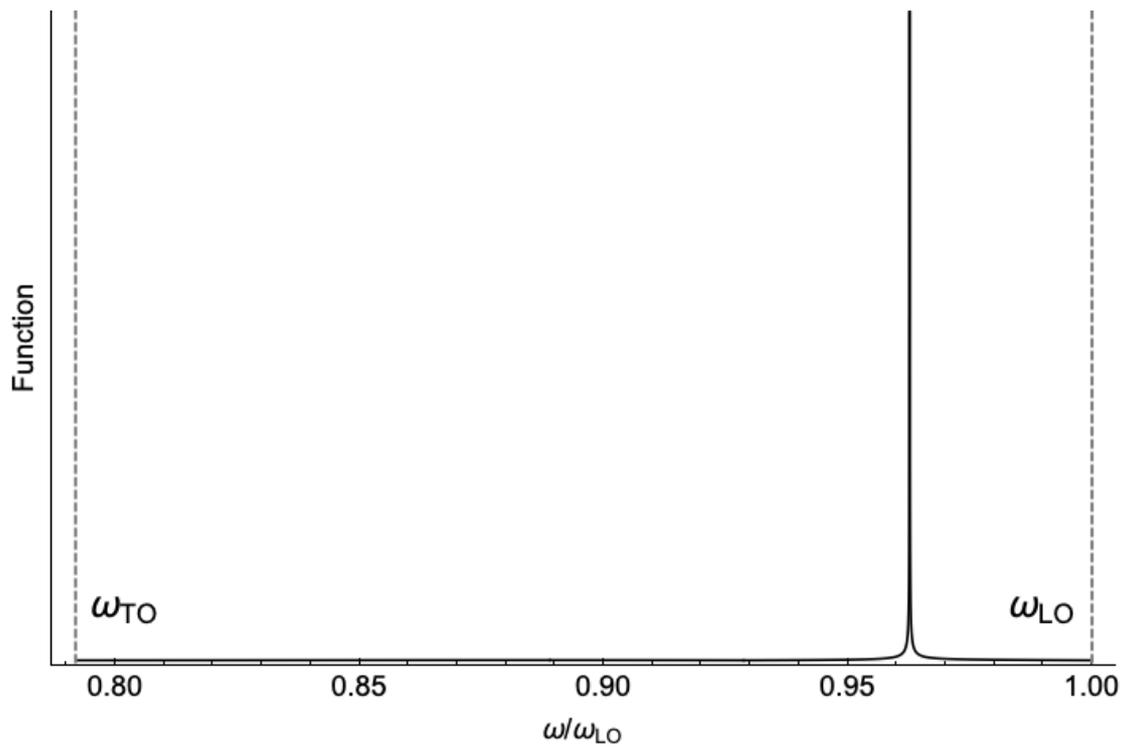

Figure 1 The integrand of Equation 9 which satisfies the disappearance of the sum of the dielectrics plotted as a function of frequency.